\documentclass[12pt]{article}

\usepackage{newtxtext,newtxmath}

\usepackage{graphicx}

\usepackage[letterpaper,margin=1in]{geometry}

\renewenvironment{abstract}
	{\quotation}
	{\endquotation}

\date{}

\makeatletter
\renewcommand{\fnum@figure}{\textbf{Figure \thefigure}}
\renewcommand{\fnum@table}{\textbf{Table \thetable}}
\makeatother

\usepackage{scicite}

\usepackage{url}

\def\scititle{
	Probing the Planck scale with quantum computation
}
\title{\bfseries \boldmath \scititle}

\author{
	Boaz~Katz$^{1\ast}$,
	Shlomi~Kotler$^{2\ast}$\\
	\small$^{1}$Department of Particle Physics and Astrophysics, Weizmann Institute of Science, Rehovot 76100, Israel.\and
	\small$^{2}$Racah Institute of Physics, The Hebrew University of Jerusalem, Jerusalem 91904, Israel.\and
	\small$^\ast$Corresponding author. Email: boaz.katz@weizmann.ac.il;~shlomi.kotler@mail.huji.ac.il\and
}

\begin{document} 

\maketitle

\begin{abstract} \bfseries \boldmath
    General relativity and quantum mechanics are incompatible at the Planck scale.‌ This contention can be examined  if a quantum computer is set to operate at a rate that exceeds the classical limit of one operation per Planck volume-time, or equivalently $2^{491}$~m$^{-3}$\ s$^{-1}$. Here we quantify the relation between the logical qubit count and the extent to which classicality is challenged. We argue that 500 logical qubits are sufficient to reject theories confined to a laboratory. We account for the operational cost of computation and communication at all scales up to and including the observable universe, ultimately constrained by a 1600-logical-qubit computer. Remarkably, current plans for commercial quantum computers are projected to surpass this limit, thereby putting the quantum-gravity standoff to the test.
\end{abstract}

\noindent
Our current understanding of the universe at large distances, embodied by the theory of general relativity, is at odds with that of the sub-atomic sizes governed by quantum mechanics. On astrophysical scales, quantum corrections to gravity are negligible, while on microscopic scales, gravity is too weak to be observed. The Planck scale is a point of contention where both should be significant and in contradiction. Addressing this problem requires devising experiments that probe the constituents of the universe at distances of $l_P\equiv \sqrt{\hbar G/c^3
}\approx 1.6\times 10^{-35}$~m, at times of $t_P\equiv l_p/c \approx 5.4\times10^{-44}$~s, or at energies of $E_P\equiv \hbar/t_P\approx 1.2\times 10^{28}$~eV,  where $c$ is the speed of light, $\hbar$ is the reduced Planck constant and $G$ is the gravitational constant. 

The large energies involved preclude the direct approach of using particle accelerators, as illustrated by the fact that even the most energetic accelerator to date, the Large Hadron Collider, reaches $\sim 10^{13}$~eV, which is 15 orders of magnitude smaller than the Planck energy~\cite{2008:evans:lhc-machine}. A promising indirect approach to explore the boundary between quantum mechanics and gravity is to search for anomalies in sensitive measurements. These include astronomical observations~\cite{2025:alves-batista:white-paper-and-roadmap}, massive quantum systems~\cite{2025:bose:massive-quantum-systems,2019:carney:tabletop-experiments-for-quantum}, and laser interferometers~\cite{2017:chou:the-holometer:-an-instrument-to-probe,2015:collaboration:advanced-ligo}. 

Quantum computation allows for new tests of the fundamental laws of nature due to its extraordinary property of achieving an exponential number of operations~\cite{1997:deutsch:the-fabric-of-reality,2016:t-hooft:the-cellular-automaton-interpretation,2026:palmer:rational-quantum-mechanics:,2005:aaronson:limits-on-efficient-computation}. In particular, the ability to reach very high computational rate densities that exceed one operation per Planck volume per Planck time may allow a direct test of classical theories at this scale~\cite{2016:t-hooft:the-cellular-automaton-interpretation}. Here we quantify the relation between computational capability and the resulting constraints on classical theories. We show that theories that are confined to a typical laboratory volume and experiment time can be ruled out by a quantum computer with approximately 500 logical qubits. We consider more extensive theories that account for computation and communication costs at ever-growing scale and should therefore be contended against larger and larger quantum computers. The most inclusive theory, describing a fully connected universe that is only limited by causality, corresponds to approximately 1600 logical qubits. As a result, we argue that Planck physics will soon be probed by quantum computers aiming at breaking RSA-2048 encryption by implementing Shor's algorithm for integer number factoring. 

\begin{figure}[h] 
		\centering
		\includegraphics[width=0.4\textwidth]{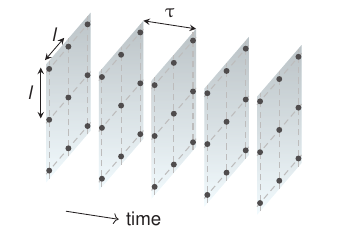} 
		\caption{\textbf{Computational Rate Density (CRD).} A simplified model of a computational process. Computing elements are spaced at a distance $l$ from one another, performing an operation (represented by dots) every clock cycle $\tau$. The resulting number of operations per unit volume per unit time (CRD) is $\mathcal{C}=1/(l^3\tau)$; see Eq.~\ref{eq:calc_density}. The spatial coordinates are represented here by a two-dimensional grid for simplicity. }
		\label{fig:lab}
\end{figure}

A simplified model of a computer consists of a grid of computing elements, at a distance $l$ from one another, that perform a single operation every time step $\tau$, as shown in Fig.~\ref{fig:lab}. For example, in a contemporary processor, the computing elements are transistors, separated by $l\sim 50$~nm and operating at a clock cycle of $\tau\sim10^{-10}$~s. Inverting this perspective, given a computer with hidden internal elements, a constraint on $l$ can be derived based on its performance. Put simply, if $N_\mathrm{ops}$ is the number of operations performed in a single cycle, then $l$ must obey $l \leq (V_3/N_\mathrm{ops})^{1/3}$, where $V_3$ is the volume of the computer. 

A more general case involves a black-box computer that has demonstrated $N_\mathrm{ops}$ operations within a time span $T$. In this case, since $\tau$ is unknown, a strict upper limit on $l$ can be derived from the fact that information cannot propagate faster than the speed of light, limiting the cycle time to $\tau \ge l/c$. Therefore the length scale must satisfy:
\begin{equation}
l \le \left(\frac{V_3cT}{N_{\mathrm{ops}}}\right)^{1/4}.
\label{eq:calc_lab}
\end{equation}
\noindent\ This bound depends on the intrinsic Computational Rate Density (CRD) of the computer,
\begin{equation}
\mathcal{C}\equiv\frac{N_{\mathrm{ops}}}{V_3T}=\frac{1}{l^3\tau},
\label{eq:calc_density}
\end{equation}
\noindent\ i.e., the number of operations per unit volume per unit time. Equation~\ref{eq:calc_lab} can be restated as $\mathcal{C}\le c/l^4$.

The upper limits obtained by the current capabilities of classical computers using Eq.~\ref{eq:calc_lab} do not add new constraints on known physics. For example, a modern GPU die, with volume $\sim 744$~mm$^3$ capable of $3352$ trillion operations per second\cite{2026:nvidia}, translates to a conservative upper limit of $l\lesssim 0.5$~mm. In fact, current technologies are ultimately limited by the atomic scale and will not be able to probe lengths smaller than $\sim 1$~\AA. 

\begin{figure} 
	\centering
	\includegraphics[width=1\textwidth]{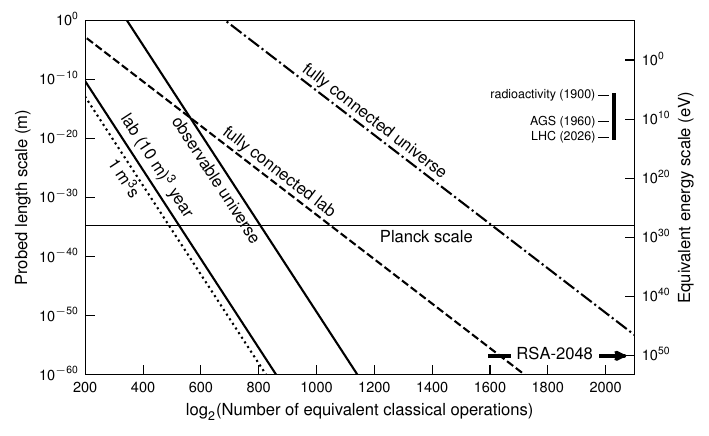} 
	\caption{\textbf{Length scale probed by a quantum computer.}
	The probed length scale is shown versus the Number of Equivalent classical Operations (NEO) demonstrated by a verified calculation of a quantum algorithm. 
    The dotted line corresponds to a small experiment of size 1~m$^3$ running for 1~s while the lower solid line corresponds to a large laboratory building of size 1000~m$^3$ running for a full year (Eq.~\ref{eq:calc_lab}). The upper solid line extends the resources of the lab to include all possible calculations within its past light cone throughout the history of the universe since the Big Bang (Eq.~\ref{eq:calc_universe}). The dashed line corresponds to a fully connected lab where each computational event integrates direct inputs from all previous calculations in its causal past (Eq.~\ref{eq:calc_lab_fully_connected} and Fig.~\ref{fig:connectivity}B). The dash-dotted line represents this fully connected computation when extended to the entire observable universe (Eq.~\ref{eq:calc_universe_fully_connected} and Fig.~\ref{fig:universe}). The range of the estimated number of logical qubits required to break a modern RSA code using Shor's algorithm is marked on the x-axis \cite{2010:nielsen:quantum-computation-and-quantum-information,2025:chevignard:reducing-the-number-of-qubits}. The y-axis on the right-hand side shows the corresponding energy scales. Marked years 1900, 1960 and 2026 correspond to the highest particle energies probed at those eras with radioactivity, the Alternating Gradient Synchrotron~\cite{1958:beth:the-Brookhaven-alternating-gradient-synchrotron}, and the Large Hadron Collider~\cite{2008:evans:lhc-machine}, respectively. }
	\label{fig:length-vs-qubits}
\end{figure}

Quantum computers dramatically increase the CRD. Specifically, a quantum computer with $n$ logical qubits is expected to perform  
\begin{equation}
N_{\mathrm{ops}}\geq 2^{n}
\end{equation}
\noindent\ equivalent classical operations. The resulting length scale probed by quantum computers is shown in Fig.~\ref{fig:length-vs-qubits} versus the logarithm of the Number of Equivalent classical Operations (NEO). This figure encapsulates the main results of this paper. As can be seen, the trend line of a large lab ($1000$~m$^3$), operating for a full year, will reach the Planck scale with $n=525$ logical qubits. Such a computer will reach the Planck computational rate density of  
\begin{equation}
\mathcal{C}_{P}\equiv \frac1{l_P^{3}t_P} \approx 1.37\times 2^{490}~\mathrm{ops~m}^{-3}\mathrm{s}^{-1}.
\end{equation}
\noindent\ Given that quantum computers are planned to accommodate a much larger number of logical qubits, their computational rate density is expected to far exceed $\mathcal{C}_{P}$, requiring any underlying classical elements to be much smaller than the Planck scale.  

We next discuss important extensions of this bound with fewer restrictions on computational power. Computers often increase their capacity by accessing additional processors using a shared communication network. Moreover, a computation may incorporate tabulated results from previous calculations. Therefore, the possible NEO of a lab may be much larger than that used in Eq.~\ref{eq:calc_lab}. Estimating its magnitude requires knowledge of the details of the network and the data storage capability of its processors. 

Regardless of the intricacies of computing systems, all are ultimately limited by causality. The latter can be used to set upper bounds on computation capacity. For an external resource to contribute, it must be in the past light cone of the final output. The portion of the light cone that needs to be accounted for should include all preceding calculations that were tabulated, requiring knowledge of their history. 

\begin{figure}[h]
    \centering
    \includegraphics{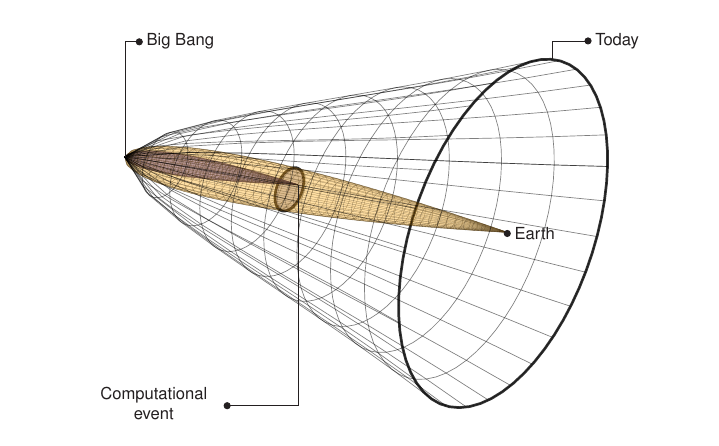}
    \caption{\textbf{Causal history of a computation on Earth today since the Big Bang.} The expansion history of the universe is shown by a black wireframe whose diameter is proportional to the cosmological scale factor $a(t)$. Any computational event within the past light cone (orange wireframe) of an experiment today may have contributed to its result and is therefore accounted for in the spacetime volume calculated in Eq.~\ref{eq:calc_universe}, corresponding to the upper solid line in Fig.~\ref{fig:length-vs-qubits}. In a fully connected universe, every intermediate computational event can directly receive information from all other events within its past light cone (purple wireframe), adding to the operation count in Eq.~\ref{eq:calc_universe_fully_connected}, corresponding to the dash-dotted line in Fig.~\ref{fig:length-vs-qubits}.}
    \label{fig:universe}
\end{figure}

The most inclusive choice for the history of a computation is to extend its origin all the way to the Big Bang. In this case, the shape of the light cone is set by the expansion history of the universe, shown in Fig.~\ref{fig:universe} (orange wire-frame). At a time $t$ in the past, the universe was smaller compared to today by the cosmic scale factor $a(t)$. Naturally, not all of the universe at that time could have contributed to a calculation today. Only those events from which light had enough time to reach us should be accounted for. The distance $d(t_1,t_2)$ light travels from time $t_1$ to time $t_2$, as measured today between the emitting and receiving galaxies (co-moving distance), is given by,
\begin{equation}
	d(t_1,t_2) = \int_{t_1}^{t_2} \frac{cdt}{a(t)}.
\label{eq:d}
\end{equation}
~\noindent\ The 3-dimensional region at $t_1$ from which calculations can affect an operation at $t_2$ is therefore a sphere of radius $a(t_1)d(t_1,t_2)$, as measured at $t_1$. Integrating all of these spheres since the Big Bang results in the total spacetime volume that can affect a calculation at $t_2$,
\begin{equation}
V_4(t_2) = \frac{4\pi}{3}\int_0^{t_2} dt_1 a^3(t_1)d^3(t_1,t_2).
\label{eq:V4}
\end{equation}
\noindent\ The number of operations available to an experiment today cannot exceed that of a universe densely packed at the upper CRD limit of $c/l^4$, 
\begin{equation}
	N_\mathrm{ops}= \frac{cV_4(T_U)}{l^4} =k_{4U}\left(\frac{c/H_0}{l}\right)^4,
	\label{eq:calc_universe} 
\end{equation}
\noindent\ where $T_U\approx 14$~Gyr is the age of the universe, $H_0\approx 70~ \rm km~s^{-1}~Mpc^{-1}$ is the Hubble constant, and $k_{4U}\approx 0.13$ is a dimensionless factor set by the cosmological parameters \cite{methods}. Shown by the upper solid line in Fig.~\ref{fig:length-vs-qubits}, this limit intersects the Planck scale at a threshold of $\log_2(N_\mathrm{ops})\approx 806$ logical qubits. Note that the implied number of operations available in the entire universe may seem larger than the estimate in Ref.~\cite{2002:lloyd:computational-capacity-of-the-universe}. Those differ, however, since the latter enumerates quantum rather than classical operations. 

\begin{figure}[h] 
    \centering
    \includegraphics[width=0.4\textwidth]{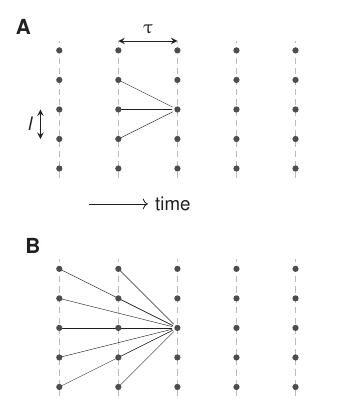}
    \caption{\textbf{Computational connectivity.} Possible communications (illustrated by lines) between computational events (illustrated by dots). The laboratory computational process in Fig.~\ref{fig:lab} is represented here with one spatial dimension. Two limiting cases of connectivity are drawn. (\textbf{A}) Nearest-neighbor connectivity. Each computational event integrates the output of its adjacent events from the previous clock cycle. (\textbf{B}) Fully connected laboratory. Each computational event incorporates all prior events that are in its past light cone. For the typical scenario depicted here, the computation extends over a time that is much longer than the lab light-crossing time. Therefore, effectively all preceding computational events from all elements contribute; see Eq.~\ref{eq:calc_lab_fully_connected}.}
    \label{fig:connectivity}
\end{figure}

Communications between computing elements increase the number of operations and should also be accounted for. In the case of nearest-neighbor connectivity shown in Fig.~\ref{fig:connectivity}A, where each event is influenced by a small number of predecessors, the increased computational overhead will have negligible impact. For example, in the laboratory scenario, eight inputs per operation will modify the required number of logical qubits needed to reach the Planck scale from 525 to 528. By contrast, networks that exhibit much higher linkage may impact the overall operation count significantly. Examples include distributed systems~\cite{1978:lamport:time-clocks-and-the-ordering} and biological neural networks. 
 
Different choices of network connectivity may result in significantly different operation counts. For a natural example of a relativistic connectivity, see Supplementary Text. All possible choices, however, can be bounded by the fully connected mesh, where all causally connected events are included. In this case each computational event integrates inputs from all events in its past light cone. In a typical laboratory, the computation time, $T$, is much longer than the lab light-crossing time. Therefore, each computational event will include inputs from almost all previous events, since only a small fraction occur within the preceding light-crossing time. Neglecting this minor correction, every pair of events is connected and should be counted once. The resulting number of operations is therefore 
\begin{equation}
	N_\mathrm{ops} = \frac12\left(\frac{V_3cT}{l^4}\right)^2.
	\label{eq:calc_lab_fully_connected} 
\end{equation}
\noindent\ Full connectivity is illustrated in Fig.~\ref{fig:connectivity}B. The length scale probed by the fully connected lab is shown by the dashed line of Fig.~\ref{fig:length-vs-qubits}. It intersects the Planck scale at 1050 qubits. 

We are now in a position to account for the largest possible extension for computation capacity\textemdash the fully connected universe. It involves full-connectivity extended to the entire universe since the Big Bang. This calculation can be broken down as follows. At each time $t$, all computational events that could influence today are encapsulated within a sphere of radius $a(t)d(t,T_U)$, shown by a dark circle in Fig.~\ref{fig:universe}. Each one of these events, in turn, can be affected by all events within its own past light cone. The corresponding spacetime volume $V_4(t)$ is depicted by the purple wireframe in Fig.~\ref{fig:universe}. The total number of operations is obtained by integrating the product of these two factors over the history of the universe:
\begin{equation}
	N_\mathrm{ops}= \frac{4\pi c^2}{3l^8}\int_0^\mathrm{T_U} dt~ a^3(t)~ d^3(t,T_U)V_4(t)=k_{8U}\left(\frac{c/H_0}{l}\right)^8,
	\label{eq:calc_universe_fully_connected} 
\end{equation}
\noindent\ where $k_{8U}\approx 8.6\times 10^{-4}$ is a second dimensionless factor set by the cosmological parameters~\cite{methods}. The resulting limit is shown in Fig.~\ref{fig:length-vs-qubits} by the dashed-doted line. 

Even for this most extensive model, the Planck scale will be probed for machines with only $1609$ logical qubits, within the requirements to break RSA-2048 encryption~\cite{2010:nielsen:quantum-computation-and-quantum-information,2025:chevignard:reducing-the-number-of-qubits}.  To appreciate the magnitude of this ultimate NEO limit, it is worth revisiting its underlying physical constituents, depicted in Fig.~\ref{fig:universe}. The universe is tightly packed with computing elements, at a Planck distance from one another, performing calculations every Planck time. Moreover, as the universe expands, new elements are constantly being added, filling the newly created gaps. Accounting for the operational cost of communication, the result of a calculation today includes direct inputs from all events in its past light cone since the Big Bang. Finally, each of these individual past events carries it own past light cone, also furnished with computational events that are densely packed at the Planck scale. 

The bounds above demonstrate that a quantum computer that successfully factorizes numbers with $n=2048$ binary digits will all but rule out models in which the universe has classical rules at the Planck scale, such as those discussed in Refs.~\cite{2012:zenil:a-computable-universe:-understanding-36,2012:zenil:a-computable-universe:-understanding-19,2016:t-hooft:the-cellular-automaton-interpretation}. 
A reservation to this conclusion is that the underlying classical evolution may have performed substantially fewer than $2^n$ operations. In fact, there are classical algorithms that can factor numbers much more efficiently~\cite{1990:lenstra:the-number-field-sieve}, and those could be further improved in the future. The existence of such algorithms, however, is not the determining factor for our purposes, unlike quantum computational advantage. Here, any contending classical explanation of the computation would not only have to account for number factoring, but also mimic the steps of the specific quantum algorithm implementation. Indeed, the process of developing quantum computers entails rigorous tests of their memory elements, quantum gates, and algorithmic submodules. We believe that the combined evidence provided by a solution to a computationally hard problem that can be verified, together with access to sub-components and interim results, would tilt the scale in favor of quantum mechanics. 

To date, there have been experimental demonstrations that involved up to a few dozen logical qubits~\cite{2024:bluvstein:logical-quantum-processor,2024:hetenyi:creating-entangled-logical,2024:paetznick:demonstration-of-logical-qubits,2025:acharya:quantum-error-correction} and there are detailed plans to extend these numbers to the thousands in order to break RSA-2048~\cite{2025:gidney:how-to-factor-2048-bit-rsa-integers,2025:chevignard:reducing-the-number-of-qubits,2025:zhou:resource-analysis-of-low-overhead,2025:yoder:tour-de-gross:-a-modular,2026:cain:shors-algorithm-is-possible}. An exciting alternative that may allow reaching high CRD sooner is to use Noisy Intermediate-Scale Quantum devices~\cite{2018:preskill:quantum-computing-in-the-nisq} that run algorithms such as boson sampling~\cite{2011:aaronson:the-computational-complexity-of-linear}. This was demonstrated, for example, with random circuits~\cite{2019:arute:quantum-supremacy-using} and Gaussian boson sampling~\cite{2025:liu:robust-quantum-computational}. While experiments with noisy systems have shown faster progress, they involve a nontrivial reduction of computational complexity. Quantifying the NEO of such experiments is a worthwhile endeavor that is beyond the scope of this paper.
 

Does quantum mechanics have boundaries? If so, what experimental axes lead there? Newtonian mechanics breaks down at high speeds. Classical physics fails at subatomic scales. There are no known analogous limits to quantum mechanics. At the Planck length, quantum mechanics clashes with another successful theory---general relativity. At this small scale, at least one of these theories must fail. Unfortunately, direct experiments show little hope of testing this regime in the foreseeable future. As a result, indirect approaches are currently being pursued. Quantum computation is an emerging technology that may soon push the boundary of experimental physics along a completely new axis---Computational Rate Density (CRD). Quantum mechanics has a unique potency to condense an exponential number of equivalent classical operations (NEO) into the volume and time span of a laboratory experiment. Remarkably, future quantum computers that are currently under industrial development are expected to far exceed the Planck CRD of $\approx 1.37\times2^{490}$~operations\,m$^{-3}$\,s$^{-1}$, eventually surpassing the computational capacity of the fully connected universe. If successful, this will vindicate quantum mechanics and challenge our current fundamental theory of gravity. If, however, quantum computation efforts persistently fail, with no apparent technical reasons, this may be the first sign of the limits of quantum mechanics. In either case, it appears that this extensive human endeavor, which is largely driven by its technological potential, may soon probe into some of the deepest mysteries of nature.



\newpage



%


\clearpage 

%
\bibliography{refs} 
\bibliographystyle{sciencemag}

\subsection*{Supplementary materials}
Materials and Methods\\
Supplementary Text\\
References \textit{(33-\arabic{enumiv})}\\ 


\newpage


\renewcommand{\thefigure}{S\arabic{figure}}
\renewcommand{\thetable}{S\arabic{table}}
\renewcommand{\theequation}{S\arabic{equation}}
\renewcommand{\thepage}{S\arabic{page}}
\setcounter{figure}{0}
\setcounter{table}{0}
\setcounter{equation}{0}
\setcounter{page}{1} 


\begin{center}
\section*{Supplementary Materials for\\ \scititle}

Boaz~Katz$^{\ast}$\\
Shlomi~Kotler$^{\ast}$\\ 
\small$^\ast$Corresponding author. Email: boaz.katz@weizmann.ac.il;~shlomi.kotler@mail.huji.ac.il\\
\end{center}

\subsubsection*{This PDF file includes:}
Materials and Methods\\
Supplementary Text\\

\newpage


\subsection*{Materials and Methods}



\subsubsection*{Detailed calculation of the cosmological pre-factors}
We adopt a flat Lambda-CDM (cold dark matter) cosmological model for the expanding universe~\cite{2020:planck-collaboration-cosmological-parameters}. For simplicity, we round the model parameters to one significant digit within their experimental uncertainty. This results in the following parameters: unitless density parameters $\Omega_M=0.3$ for matter, $\Omega_\Lambda=0.7$ for dark energy and a Hubble constant value of $H_0=70$~km\,s$^{-1}$\,Mpc$^{-1}$ \cite{2020:planck-collaboration-cosmological-parameters,2022:riess:a-comprehensive-measurement-of-the-local}. We neglect the contribution of radiation density ($\Omega_\mathrm{rad}\sim 10^{-4}$). The scale factor $a(t)$ is given by,
\begin{equation}
a(t)=\left(\frac{\Omega_M}{\Omega_\Lambda}\right)^{1/3} \sinh^{2/3}(t/t_\Lambda),
\end{equation}
\noindent\ where $t_\Lambda=2/(3H_0\sqrt{\Omega_\Lambda})$.
The age of the universe for these parameters is $T_U=13.5$ Gyr. 

The resulting numerical constants appearing in Eq.~\ref{eq:calc_universe} and Eq.~\ref{eq:calc_universe_fully_connected} of the main text are, 
\begin{equation}
	k_{4U}=\frac{H_0^4}{c^3}V_4(T_U)=0.13,
\end{equation}
\noindent\ and,
\begin{equation}
	k_{8U}=\frac{4\pi H_0^8}{3c^6} \int_0^{T_U} dt V_4(t) a^3(t)d^3(t,T_U)=8.6\times 10^{-4},
\end{equation}
\noindent\ respectively, where we used Eq.~\ref{eq:d} and Eq.~\ref{eq:V4} of the main text. 


\subsection*{Supplementary Text}
\subsubsection*{Example of relativistic connectivity}

A simple model of connectivity that is manifestly relativistic involves a distributed system of moving computing elements~\cite{1978:lamport:time-clocks-and-the-ordering} that perform calculations with proper time durations $\tau$ and broadcast their results at the end of each calculation. Signals are transmitted to all directions at the speed of light with no loss of information. Each element, in turn, integrates the signals it received to an extent that depends on the connectivity. In this model, the data from a signal that was received by an element may be used in many following calculations and each such usage is counted as an additional operation. 

The computational spacetime events are associated with the locations and times of broadcasts. These events are equally spaced along the worldlines of the computing elements with equal proper time spacing $\tau$. The output of a computational event $A_1$ by element $A$ will be considered as an input for an event $B_1$ by element $B$ if the broadcast from $A_1$ was received by $B$ prior to $B_1$ and integrated in the calculation that lead to $B_1$. Full connectivity is achieved if elements use in each calculation all previous signals they received. This implies that each computational event uses as direct inputs the results of all previous events in its past light cone as described in the main text.

A simple choice for a limited connectivity within this model is to include in a computational event only signals that were received by the computing element during the time interval $\tau$ that preceded the event. This significantly reduces the memory requirements of the model. We next estimate the resulting number of operations of a computer for this choice for the two cases of resource accessibility considered in the main text: first, where communication is limited to the lab, and second, where it is unlimited and the lab can access all previous calculations in the observable universe.

For communications that are limited to the lab, the duration of the entire calculation is typically much larger than the light-crossing time of the computer. In this case, every computing element will receive one broadcast per time step from all the other $V_3/l^3$ elements. The resulting number of operations is therefore,
\begin{equation}
	N_\mathrm{ops} = \left(\frac{V_3}{l^3}\right)^2\frac{T}{\tau}.
	\label{eq:calc_lab_nearest_neighbor} 
\end{equation}
\noindent\ In the Planck limit of $l=l_P$ and $\tau=l/c=t_P$, this corresponds to $882$ logical qubits. Even for this limited connectivity, the resulting threshold is much higher compared to the laboratory bound of $525$ that was obtained in the main text by ignoring the impact of communication. 

We next estimate the number of operations obtained within this model if the communication spans the observable universe. As in the main text, we assume that the universe is filled with co-moving computing elements that are spaced by $l$, with new elements constantly being added as the universe expands. The rate density of computational events is thus $\mathcal{C}=1/(l^3\tau)$ everywhere. 
The number of accumulated signals that arrive at a given element up to time $t$ is given by $\mathcal{C}V_4(t)$. The rate at which the signals arrive is thus $\mathcal{C}\dot V_4$, where $\dot V_4=dV_4(t)/dt$, so that each calculation integrates $\mathcal{C}\dot V_4\tau$ inputs. The total amount of communication operations that can influence an output today is therefore,
\begin{equation}
N_{\mathrm{ops}}=\frac{4\pi\mathcal{C}}{3}\int_0^{T_U} \mathcal{C}\dot V_4\tau~ a(t)^3d^3(t,T_U)dt.
\label{eq:calc_U_special}
\end{equation}

\noindent\ An explicit expression for $\dot V_4$ can be obtained using equations Eq.~\ref{eq:d} and Eq.~\ref{eq:V4} of the main text,
\begin{equation}
\dot V_4(t_2)= 
4\pi~\int_0^{t_2}a^2(t_1)d^2(t_1,t_2)~\frac{a(t_1)}{a(t_2)}~c dt_1.
\label{eq:dotV4}
\end{equation}
\noindent\ The resulting arrival rate of signals at $t_2$, $\mathcal{C}\dot V_4(t_2)$, takes the form of a sum of contributions from different distances, $a(t_1)d(t_1,t_2)$, and their corresponding emission times, $t_1$. The broadcast rate from each of these spherical shells is $\mathcal{C}\cdot 4\pi a^2(t_1)d^2(t_1,t_2)cdt_1$ and is red-shifted by $a(t_1)/a(t_2)$.
Setting $\tau$ at the causal limit, $\tau=l/c$, the resulting total number of operations using equations Eq.~\ref{eq:calc_U_special} and Eq.~\ref{eq:dotV4} can be expressed as, 
\begin{equation}
N_{\mathrm{ops}}=k_{7U}\left(\frac{c/H_0}{l}\right)^7,
\end{equation}
\noindent\ where, 
\begin{equation}
k_{7U}=\frac{4\pi H_0^7}{3c^6}\int_0^{T_U}  a^3(t)d^3(t,T_U)\dot V_4(t)dt, 
\end{equation}
\noindent\ is a third dimensionless parameter that depends on the cosmological parameters and is approximately equal to $k_{7U}=6.2\times 10^{-3}$ for the choice of values described in~\cite{methods}.



\end{document}